# Experimental comparison of autodyne and heterodyne laser interferometry using a Nd:YVO$_4$ microchip laser

**Olivier Jacquin, Eric Lacot,[*] Wilfried Glastre, Olivier Hugon,**

**and Hugues Guillet de Chatellus**

*Centre National de la Recherche Scientifique / Université de Grenoble 1,*

*Laboratoire Interdisciplinaire de Physique, UMR 5588,*

*Grenoble, F- 38041, France*

[*]*Corresponding author: eric.lacot@ujf-grenoble.fr*

Using a Nd:YVO$_4$ microchip laser with a relaxation frequency in the megahertz range, we have experimentally compared a heterodyne interferometer based on a Michelson configuration with an autodyne interferometer based on the laser optical feedback imaging (LOFI) method regarding their signal to noise ratios. In the heterodyne configuration, the beating between the reference beam and the signal beam is realized outside the laser cavity while in the autodyne configuration, the wave beating takes place inside the laser cavity and the relaxation oscillations of the laser intensity then play an important part. For a given laser output power, object under investigation and detection noise level, we have determined the amplification gain of the LOFI interferometer compared to the heterodyne interferometer. LOFI interferometry is demonstrated to show higher performances than heterodyne interferometry for a wide range of laser power and detection level of noise. The experimental results are in good agreement with the theoretical predictions. © *2011 Optical Society of America*   OCIS   codes:   *110.3175,   280.3420.*

# 1. INTRODUCTION

When a frequency shift is introduced between the two beams in an interferometer, one realizes the so-called heterodyne interferometry. Resulting from this shift, the interference between the two waves produces an intensity modulation, at the beat frequency, which can be measured by a photodetector. In this paper, we refer to autodyne laser interferometry when the heterodyne wave mixing takes place inside the cavity of the laser source, while we speak about heterodyne laser interferometry when the mixing is realized directly on the photodetector (i.e. outside the laser cavity).

Since the development of the first laser in 1960, laser heterodyne interferometry has become a useful technique on which many high accuracy measurement systems for scientific and industrial applications are based [1]. Since the pioneer work of K. Otsuka, on self-mixing modulation effects in class-B laser [2] the sensitivity of laser dynamics to frequency-shifted optical feedback has been used in autodyne interferometry and metrology [3], for example in self-mixing laser Doppler velocimetry [4-7], vibrometry [8-10], near field microscopy [11,12] and laser optical feedback imaging (LOFI) experiments [13-15]. Compared to conventional optical heterodyne detection, frequency-shifted optical feedback shows an intensity modulation contrast higher by several orders of magnitude and the maximum of the modulation is obtained when the shift frequency is resonant with the laser relaxation oscillation frequency [4, 16]. In this condition, an optical feedback level as low as -170 dB (i.e. $10^{17}$ times weaker than the intracavity power) has been detected [5].

In previous papers [16, 17], we have demonstrated that in autodyne interferometry, the main advantage of the resonant gain (defined by the ratio between the cavity damping rate and the population-inversion damping rate of the laser) is to raise the laser quantum noise over the detection noise in a relatively large frequency range close to the laser relaxation frequency. Under these conditions, the signal to noise ratio (SNR) of a LOFI setup is frequency independent and, more importantly, shot noise limited. We have also theoretically established that for the dynamical range of a LOFI setup to be maximized, the best value of the shift frequency is not the relaxation frequency, but the frequency at which the amplified laser quantum noise is equal to the detection noise level. At this particular frequency, the value of the LOFI gain is simply given by the ratio between the power density levels of the detection noise and of the shot noise.

In the same paper [17] we have also theoretically compared the SNR of a LOFI setup (autodyne interferometer) with the SNR of a conventional Michelson configuration (heterodyne interferometer) and we have found that, irrespective to the laser dynamical parameters, the LOFI SNR gain (i.e. the ratio of the LOFI SNR to the Michelson SNR) is still given by the ratio between the power density levels of the detection noise and of the shot noise.

The main objective of the present study is to experimentally verify the theoretical predictions mentioned in [17]. The present paper is organized as follows: in the first section, we give a basic description of the two types of experimental setups (autodyne and heterodyne interferometers). In the second section, we present a theoretical laser model based on quantum Langevin equations [18,19] to fit the experimental noise power spectrum of our Nd:YVO$_4$ microchip laser. The LOFI signal gain and the LOFI SNR gain of the LOFI setup can be predicted from this fit. In the third section, we measure directly the SNR of the heterodyne and the LOFI setups. The LOFI SNR gain (i.e. the ratio of the LOFI SNR to the Michelson SNR) is then experimentally determined and compared with the theoretical prediction obtained from the study of the laser noise power spectrum. Finally we conclude on the advantages and disadvantages of the LOFI method compared to an heterodyne interferometer in terms of SNR and practical use.

## 2. STUDIED INTERFEROMETRIC SETUPS

A schematic diagram of the LOFI experimental setup is shown in Fig. 1(a). Typically, the laser is a CW microchip laser [20]. The laser beam is conventionally sent towards a frequency shifter that is composed of two acousto-optic deflectors (AOD), respectively supplied by a RF signal at 81.5 MHz and 81.5 MHz+$F_e$/2 where $F_e$ is a tunable RF frequency. The AODs are arranged so that the laser beam is successively diffracted in the -1 order of the first AOD and the +1 order of the second AOD. At this stage, the resulting optical frequency shift of the laser beam is $F_e$/2. Next the beam is sent onto the target using a lens and galvanometric scanner made by two rotating mirrors. A part of the light diffracted and/or scattered by the target is then re-injected inside the laser cavity after a second pass in the frequency shifter. Therefore, the optical frequency of the reinjected light is shifted by $F_e$ [21]. This frequency can be adjusted and is typically of the order of the laser relaxation frequency $F_R$. From the geometrical point of view, the laser beam waist and the laser focal spot on the target under investigation are optically conjugated through the lenses $L_1$ and $L_2$. The amount of optical feedback is characterized by the effective power reflectivity ($R_e$) of the target, where $R_e$ takes into account the target albedo, the numerical aperture of the collecting optics, the transmission coefficients of all the optical components (except for the beam splitter and the AOD which are addressed separately) and the overlap of the retro-diffused field with the Gaussian cavity beam. In the case of a weak optical feedback, the coherent interaction (beating) between the lasing electric field and the frequency-shifted reinjected field leads to a modulation of the laser output power. For the detection purpose, a fraction of the output beam of the microchip laser is sent to a photodiode by means of a beam splitter characterized by a power reflectivity $R_{bs}$. The photodiode is assumed to have a quantum efficiency of 100%. The voltage delivered by the photodiode is finally processed by a lock-in amplifier which gives the LOFI signal (i.e. the magnitude and the phase of the retro-diffused electric field) at the demodulation frequency $F_e$. Usually, the LOFI technique is used to realize images obtained pixel by pixel (i.e. point by point, line after line) by full 2D galvanometric scanning. In the present comparative study, no scanning occurs (i.e. $R_e$ has a constant value) and the amount of optical feedback is tuned by changing the frequency shifter efficiency ($\rho$) and therefore the amount of photons sent on the target.

A schematic diagram of the heterodyne experimental setup is shown on Fig. 1(b). Compared to Fig. 1(a), the only differences are: the Faraday optical isolator which prevents any optical feedback in the laser source, the beam splitter orientation and the reference mirror (RM) which allows the mixing of the reference wave with the signal wave directly on the detector. In this configuration, the delivered voltage is also processed by a lock-in amplifier. The optical isolation is of the order of -50dB.

Thus, for a given laser output power, target and photodiode noise level, the experimental setups shown on Figs. 1(a) and 1(b) enable a direct comparison of the sensitivity of a heterodyne interferometer based on a Michelson configuration and of an autodyne interferometer based on the LOFI method. Besides, it can already be noticed that compared to the heterodyne setup, the autodyne setup does not require any delicate alignment. More precisely, the LOFI setup is even always self-aligned because the laser simultaneously fulfills the function of the source (i.e. the emitter) and of the photodetector (i.e. the receptor).

## 3. POWER SPECTRUM OF THE LASER SOURCE

### *A. Theoretical model*

For a theoretical description of the solid-state laser noise properties, we have used a full quantum model based on the Langevin equations approach [18, 19]. The model deals with a system of homogeneously broadened four-levels atoms, reduced to a three-levels laser by an adiabatic elimination of the upper level of the pumping transition. This model also assumes that the lower level of the laser transition is not the ground state and that the atoms fill the laser cavity. This model which is quite general is well adapted to describe the behavior of a Nd:YVO$_4$ microchip laser [19]. More specifically, this model allows to take into account the lifetime of the lower level of the laser transition which is usually neglected compared to the value of the laser cavity lifetime in conventional neodymium laser with decimetric or metric cavity length.

In the present study, the laser used is a Nd:YVO$_4$ microchip laser with a conventional pumping. This laser belongs to the so-called class-B lasers family [22] which is characterized by a decay rate of the atomic polarization much faster than the other relaxation rates. Therefore, according to [18,19] and setting $c \to +\infty$ (adiabatic elimination of the atomic polarization) and $s(\tilde{\Omega}) = 1$ (Poissonian pumping), the normalized intensity noise spectrum at the laser output is given by:

$$V_{out,B}(\tilde{\Omega}) = 1 + \eta \frac{2b(a+a')}{b-a'} \frac{1}{D_B(\tilde{\Omega})} \left( \begin{array}{l} [b^2 + \tilde{\Omega}^2][(a+a')^2 + \tilde{\Omega}^2]\left[\dfrac{n}{a+a'}\right] \\[1em] + 2w_B^2 \left\{ \begin{array}{l} [(b-a')^2 + \tilde{\Omega}^2]n \\[0.5em] -[(b-a')(a+a') + \tilde{\Omega}^2]\left[r - \dfrac{a+2a'}{a+a'}n\right] \\[0.5em] +[(a+a')^2 + \tilde{\Omega}^2]\left[\dfrac{a'}{a+a'}\right] \end{array} \right\} \end{array} \right) \quad (1a)$$

with the following shorthands:

$$D_B(\tilde{\Omega}) = \left| -i\tilde{\Omega}(b - i\tilde{\Omega})(a + a' - i\tilde{\Omega}) + 2w_B^2(a + b - i2\tilde{\Omega})(1 - i\tilde{\Omega}) \right|^2 \tag{1b}$$

$$n = \frac{ra + b + a'(r-1)}{a + b} \tag{1c}$$

$$w_B^2 = \frac{(a + a')b}{2(a + b)}(r - 1). \tag{1d}$$

For more consistencies, we have adopted, in the set of equations (1), the notation used in [19]: $\eta = \frac{\kappa_{out}}{\kappa}$ represents the correction for internal optical losses, $\kappa = \kappa_{out} + \kappa_{losses}$ is the total cavity damping constant, $\kappa_{out}$ is the output coupling constant and $\kappa_{losses}$ represents the internal optical losses.

The dimensionless parameters a, a', b and the dimensionless noise frequency $\tilde{\Omega}$ are defined as follows: $a = \frac{\gamma_a}{\kappa}$ (respectively $b = \frac{\gamma_b}{\kappa}$) is the normalized decay rate of the atomic population between the upper level (respectively the lower level) of the lasing transition and the ground state, while $a' = \frac{\gamma_{a'}}{\kappa}$ is the normalized decay rate of the atomic population between the two levels of the lasing transition and $\tilde{\Omega} = \frac{\Omega}{\kappa}$.

The normalized pump parameter r is defined as the ratio between the pump power $p_{pump}$ and the threshold pump power $p_{th}$: $r = \frac{p_{pump}}{p_{th}}$.

Fig. 2 shows, in a logarithmic scale, the normalized noise power spectrum of a laser calculated from the complete set of Eqs. (1). In this figure, the 0 dB level corresponds to the shot noise level. For a class-B laser ($1 \gg a, a'$), Fig. 2 shows that the laser noise power spectrum is dominated by the laser relaxation frequency ($\Omega_R$) and that the decay rate of the lower level of the laser transition ($\gamma_b$) allows to modify the resonant width ($\Delta\Omega_R$) simultaneously with the resonance height ($V_{out,B}(\Omega_R)$). For a Nd microchip laser the value of $\gamma_b$, which is affected by a huge uncertainty in the literature [19, 23], can therefore be adjusted to ensure an optimal fit of the experimental curve.

The noise power spectrum given by Eqs. (1) of the present work allows us to generalize the definition of the so called LOFI signal gain $G$ by putting [17]:

$$G(\tilde{\Omega}) = \sqrt{\frac{V_{out,B}(\tilde{\Omega}) - 1}{\eta}} . \qquad (2)$$

In order to help for the physical insight of Eq. (2), we can, for the specific case of a Nd:YVO$_4$ laser, make the following approximations: $b \gg \sqrt{w_B^2} \gg a, a'$ [19]. In such a case (corresponding to the situation shown in Fig. 2), and if we come back to physical parameters, one obtains a simplified expression:

$$G(\Omega) \approx \kappa \frac{\sqrt{(\gamma_a + \gamma_{a'})^2 r + \Omega^2}}{\sqrt{(\Omega_R^2 - \Omega^2)^2 + \Delta\Omega_R^2 \Omega^2}} \qquad (3)$$

with $\Omega_R = \sqrt{\kappa(\gamma_a + \gamma_{a'})(r-1)}$ and $\Delta\Omega_R = (\gamma_a + \gamma_{a'})\left[r + (r-1)\frac{\kappa}{\gamma_b}\right]$.

In agreement with the results of Fig. 2, Eq. (3) clearly shows that the laser noise power spectrum is dominated by the laser relaxation frequency ($\Omega_R$) and that the decay rate of the lower level of the laser transition ($\gamma_b$) allows to modify the resonant width ($\Delta\Omega_R$). At this point, one can also notice that for $b = \frac{\gamma_b}{\kappa} \to +\infty$, one retrieves $\Delta\Omega_R = (\gamma_a + \gamma_{a'})r$, which is the conventional result given in [17].

## B. Nd:YVO$_4$ microchip laser

For the experimental comparison of the autodyne and the heterodyne interferometers, we have used a Nd:YVO$_4$ microchip laser with a relatively long cavity length of $L \approx 1\,\text{mm}$ (i.e. a relatively low value of the cavity damping rate $\kappa$) by comparison with usual microchip lasers [7, 19, 20]. This laser which also has a larger atomic damping rate value $(\gamma_{a'} + \gamma_a)$ than a Nd:YAG laser [24, 25], has been chosen to limit the LOFI signal gain ($G(\Omega_R) \propto \dfrac{\kappa}{\gamma_a + \gamma_{a'}}$) and therefore to prevent any saturation effect in the autodyne interferometer [17]. With this laser, an easier analysis of the autodyne and of the heterodyne interferometers is then possible, by a direct comparison of the signal and the SNR for the same amount of light coming back from the target.

To characterize our Nd:YVO$_4$ microchip laser, we have first determined the total cavity damping constant $\kappa$ by studying the dependence of the laser relaxation oscillation frequency $(\Omega_R / 2\pi)$ on the normalized pump parameter $r$, according to the theoretical expression:

$$\Omega_R = \sqrt{\kappa(\gamma_a + \gamma_{a'})(r - 1)}. \qquad (4)$$

According to the experimental results (Fig. 3), taking $\gamma_a = 3.3 \times 10^4\,\text{s}^{-1}$ and $\gamma_{a'} = 3.3 \times 10^3\,\text{s}^{-1}$ [19, 23-25], one finds: $\kappa = 2.1 \times 10^9\,\text{s}^{-1}$.

Second, for a normalized pump parameter $r = 2.7$ (corresponding to a laser output power of $P_{out} = 40\,\text{mW}$, (i.e. $p_{out} = 2.14 \times 10^{17}\,\text{photons}/\text{s}$ at $\lambda = 1064\,\text{nm}$) and a laser relaxation oscillation frequency $F_R = 1.8\,\text{MHz}$, we have adjusted the experimental noise power spectrum of our Nd:YVO$_4$ microchip laser with the theoretical prediction, given by Eq. (1). The result is shown on Fig. 4. The parameters which ensure an optimal fit of the experimental curve are:

$\gamma_b = 3 \times 10^9 \, s^{-1}$ and $\eta = 0.3$. With those parameters, the theoretical and experimental noise power spectra are in good agreement in the vicinity of the laser relaxation frequency ($F_R$).

At low frequencies the differences between the two curves come from technical noise and/or cross saturation dynamics between modes [26-28]. At the intermediate frequencies, the harmonic noise (2xF$_R$) induced by the laser non-linear dynamics are not included in the linear analytical development used to obtain Eqs. (1) [18, 19]. At very high frequency, the level discrepancy between the two curves comes from the detection noise which is not taken into account in Eqs. (1). In Fig.4, the noise power level of the detection system is also plotted (horizontal dot line). For a detection bandwidth of $\Delta F = 500 \, Hz$, this white-noise corresponds to a Noise Equivalent Power (NEP) of $1.25 \times 10^{-9} \, W/\sqrt{Hz}$. The intersection between the detection white-noise and the laser resonant quantum-noise allows us to determine the intermediate frequency: $F_+ = 5.86 \, MHz$. Roughly speaking, for $F_e < F_+$, the sensitivity of the autodyne and the heterodyne interferometers is limited by the laser quantum noise while for $F_e > F_+$, the two interferometers are limited by the detection system noise. This last case is shown in Fig. 4.

Knowing all the laser dynamical parameters ($\kappa, \gamma_a, \gamma_{a'}, \gamma_b, r, \eta$) and using Eqs. (1) and (2), the LOFI signal gain for the resonance frequency ($F_R$) and for the intermediate frequency ($F_+$) can be estimated:

$$G(F_R) \times \eta = 7.5 \times 10^3 \qquad (5a)$$

$$G(F_+) \times \eta = 21 \qquad (5b)$$

As expected (to avoid saturation effects), the maximum value of the LOFI signal gain [Eq. (5a)] of our long Nd:YVO$_4$ microchip laser is relatively low compared to the values conventionally obtained with microchip lasers which could be of the order of $10^5 - 10^6$ [4,5, 16,19,27].

Eq. (5b) gives us the value of the LOFI SNR gain (i.e. the ratio of the autodyne SNR to the hetrodyne SNR). In agreement with Ref. [17], this value simply corresponds to the ratio between the power density spectra of the detection noise level and the shot noise:

$$\sqrt{\frac{\eta}{2 R_{bs}}} \frac{1.25 \times 10^{-9} \text{ W } \sqrt{\text{Hz}}^{-1}}{\sqrt{p_{out}}} \bigg/ \frac{hc}{\lambda} = 21,$$

where $h$ is the Planck constant and $c$ the velocity of light.

## 4. HETERODYNE INTERFEROMETER VERSUS LOFI INTERFEROMETER

The main objective of this section is to experimentally verify the predictions of the LOFI signal gain [Eq. (5a)] and of the LOFI SNR gain [Eq. (5b)] obtained from the analysis of the amplified quantum noise of our Nd:YVO$_4$ microchip laser. To do so, we have experimentally compared the signal level and the SNR of an heterodyne interferometer with the ones of a LOFI interferometer.

Figure 5 shows in a logarithmic scale, the evolution of the autodyne signal $S_{LOFI}(p_{target}, F_e)$ and of the heterodyne signal $S_{Heterodyne}(p_{target})$ versus the amount of photons sent on the target ($p_{target}$), for different values of the shift frequency ($F_e$), with: $p_{target} = (1 - R_{bs}) \rho \, p_{out}$. One can see that the LOFI signal as well as the heterodyne signal evolve linearly with $p_{t\arg et}$. One can also notice that the LOFI (autodyne) signal which is frequency dependant is higher than the Michelson (heterodyne) signal which is roughly frequency independent.

As mentioned in [17], the ratio between the autodyne signal and the heterodyne signal does not depend on the amount of light coming back from the target (i.e on $p_{t\arg et}$) and is simply given by the LOFI signal gain:

$$\frac{S_{LOFI}(p_{t\arg et}, F_e)}{S_{Heterodyne}(p_{t\arg et})} = G(F_e) \times \eta \tag{6}$$

Table 1 shows that the experimental values of the LOFI signal gain ($\frac{S_{LOFI}(p_{t\arg et}, F_e)}{S_{Heterodyne}(p_{t\arg et})}$) and the theoretical predictions $G(F_e) \times \eta$ calculated with our laser parameters are in relatively good agreement if we take into account the standard deviation of the different values.

The simultaneous analysis of the results in Figs. 4 and 5 allows us to experimentally determine the SNR of both interferometers and to compare it with the theoretical prediction. At this stage, we recall that the resonant amplification gain present in the LOFI signal $S_{LOFI}(p_{t\arg et}, F_e)$, is also present in the LOFI noise $N_{Laser}(F_e, \Delta F)$, and for $F_e < F_+$, the SNR of the LOFI setup does not depends directly of the LOFI gain $G(F_e)$ and is shot noise limited[17,27] :

$$\frac{S_{LOFI}(P_{t\arg et}, F_e)}{N_{Laser}(F_e, \Delta F)} = \frac{\sqrt{R_{bs}\eta}}{\sqrt{2\Delta F}} \sqrt{R_e}(1 - R_{bs})\rho\sqrt{p_{out}} = \frac{\sqrt{R_{bs}\eta}}{\sqrt{2\Delta F}} \frac{p_{t\arg et}\sqrt{R_e}}{\sqrt{p_{out}}}, \tag{7}$$

where the photons output rate is given by: $p_{out} = \kappa_{out} \frac{\gamma_b}{\gamma_b + \gamma_a} \frac{\gamma_{a'} + \gamma_a}{\sigma c} V(r-1)$ [19], with V the volume of the cavity mode and $\sigma$ the stimulated cross section.

On the other side, for $F_e > F_+$, the SNR of the heterodyne interferometer is also frequency independent and given by [17]:

$$\frac{S_{Heterodyne}(p_{target})}{N_{Detection}(\Delta F)} = \frac{R_{bs}\sqrt{R_e}\, p_{target}}{\left(\frac{1.25 \times 10^{-9}\, W/\sqrt{Hz}}{hc/\lambda}\right)\sqrt{\Delta F}} \qquad (8)$$

Finally, the comparison of Eqs. (7) and (8) allows us to determine the real gain (i.e. the SNR gain) of the LOFI interferometer compared to the Heterodyne interferometer:

$$\frac{S_{LOFI}(p_{target}, F_e)}{N_{Laser}(F_e, \Delta F)} \bigg/ \frac{S_{Heterodyne}(p_{target})}{N_{Detection}(\Delta F)} = \sqrt{\frac{\eta}{2R_{bs}}} \frac{\left[\frac{1.25 \times 10^{-9}\, W/\sqrt{Hz}}{hc/\lambda}\right]}{\sqrt{\langle p_{out}\rangle}} = G(F_+) \times \eta . \qquad (9)$$

Starting from the LOFI SNR and the Michelson SNR obtained from the analysis of Figs. 4 and 5, the Table 2 shows that the experimental values of the LOFI SNR gain ($\frac{S_{LOFI}(p_{target}, F_e)}{N_{Laser}(F_e, \Delta F)} \bigg/ \frac{S_{Heterodyne}(p_{target})}{N_{Detector}(\Delta F)}$) and the theoretical prediction $G(F_+) \times \eta$ are in relatively good agreement if we take into account the standard deviation of the different values. One can also observe that this gain which is the real gain of a LOFI setup is lower than the value of the resonant LOFI signal gain estimated by Eq. (6a).

Using Eq. (9), one can compare the autodyne and the heterodyne interferometers in a more general way. Indeed, for a detection noise level of $1.25 \times 10^{-9}$ W$/\sqrt{\text{Hz}}$, we obtain the same value of the SNR (i.e. $G(F_+) \times \eta = 1$) for the two kinds of interferometers, if the laser output power is increased up to 18 W (i.e. $\langle p_{out} \rangle = 9.4 \times 10^{19}$ photons/s when working at $\lambda = 1064$ nm). Likewise, for a laser output power $P_{out} = 40$ mW, one obtains $G(F_+) \times \eta = 1$, if the detection noise level is decreased down to $6 \times 10^{-11}$ W$/\sqrt{\text{Hz}}$. Finally, we can conclude that, compared to a heterodyne interferometer, the LOFI detection setup is competitive ($G(F_+) \times \eta > 1$) when working with a low power level laser and/or with a conventional noisy detection.

## 5. CONCLUSION

For a given laser output power, target under investigation and detection noise level, we have experimentally compared the SNR of a LOFI setup (autodyne interferometer) with a conventional Michelson arrangement (heterodyne interferometer). The experimental results obtained are in good agreement with the theoretical predictions given in [17].

Irrespective of the resonant value of LOFI signal gain, we have experimentally demonstrated the enhanced performances of an autodyne setup compared to a heterodyne setup thanks to a LOFI SNR gain (i.e. the ratio of the LOFI SNR to the Michelson SNR) simply given by the ratio between the power density levels of the detection noise and of the shot noise. From this study, we have concluded that the LOFI setup is competitive when the optimum value of the SNR gain is greater than unity, that is to say when working at a low laser power level, and/or with a conventional noisy detection, as described in various self–mixing metrology experiments of the review paper [3].

Finally, one can also recall that, compared to the Michelson heterodyne interferometer, the LOFI setup is always self-aligned and therefore is much more easy to implement and robust because it doesn't require any delicate alignment.

**FIGURE CAPTIONS**

Fig. 1. Schematic diagrams of the autodyne interferometer setup (a) and of the heterodyne interferometer setup (b) for scanning microscopy. $L_1$, $L_2$ and $L_3$: Lenses, OI: Optical Isolator BS: Beam Splitter with a power reflectivity $R_{bs}$, GS: Galvanometric Scanner, RM: Reference Mirror with a unitary power reflectivity $R_{rm}=1$, FS Frequency Shifter with a round trip frequency-shift $F_e$, PD: Photodiode with a white noise spectrum. The lock-in amplifier is characterized by a bandwidth $\Delta F$ around the reference frequency $F_e$. The laser output power is characterized by $p_{out}$ (photons/s), the target is characterized by its effective reflectivity $R_e \ll 1$.

Fig. 2. Calculated normalized noise power spectra of the laser output power for two different values of the dimensionless parameter: $b = \gamma_b/\kappa$. The others laser dynamical parameters correspond to a conventional $Nd^{3+}$:$YVO_4$ microchip laser: $a = 2 \times 10^{-6}$, $a' = 2 \times 10^{-7}$, $\eta = 1/2$, $r = 5$ [19]. The 0 dB level corresponds to the shot noise level.

Fig. 3. Laser relaxation frequency ($F_R = \Omega_R/2\pi$) versus the normalized pumping parameter (r). Laser dynamical parameters: $\gamma_a = 3.3 \times 10^4 \text{ s}^{-1}$ $\gamma_{a'} = 3.3 \times 10^3 \text{ s}^{-1}$ and $\kappa = 2.1 \times 10^9 \text{ s}^{-1}$.

Fig. 4. Experimental (solid-line) and theoretical (dash-line) power spectra of the laser output power of a Nd:YVO$_4$ microchip laser with: $p_{out} = 2.14 \times 10^{17}$ photons/s ($P_{out} = 40$ mW at $\lambda = 1064$ nm), $r = 2.6$, $\kappa = 2.1 \times 10^9 \text{ s}^{-1}$, $\eta = 0.3$, $\gamma_a = 3.3 \times 10^4 \text{ s}^{-1}$, $\gamma_{a'} = 3.3 \times 10^3 \text{ s}^{-1}$, $\gamma_b = 3.3 \times 10^9 \text{ s}^{-1}$. The experimental conditions are: $\Delta F = 500$ Hz; $R_{bs} = 1/10$; $R_{load} = 50$ ohm. The horizontal dot line is the detection noise level.

Fig. 5. Autodyne signal (dash-lines) and heterodyne signal (solid-lines) versus the laser power sent on the target ($p_{target}$), for different values of the shift frequency: (■) $F_e = 2.9 \times 10^6$ Hz, (▲) $F_e = 4.9 \times 10^6$ Hz, (◆) $F_e = 8.9 \times 10^6$ Hz, (▼) $F_e = 18.9 \times 10^6$ Hz.

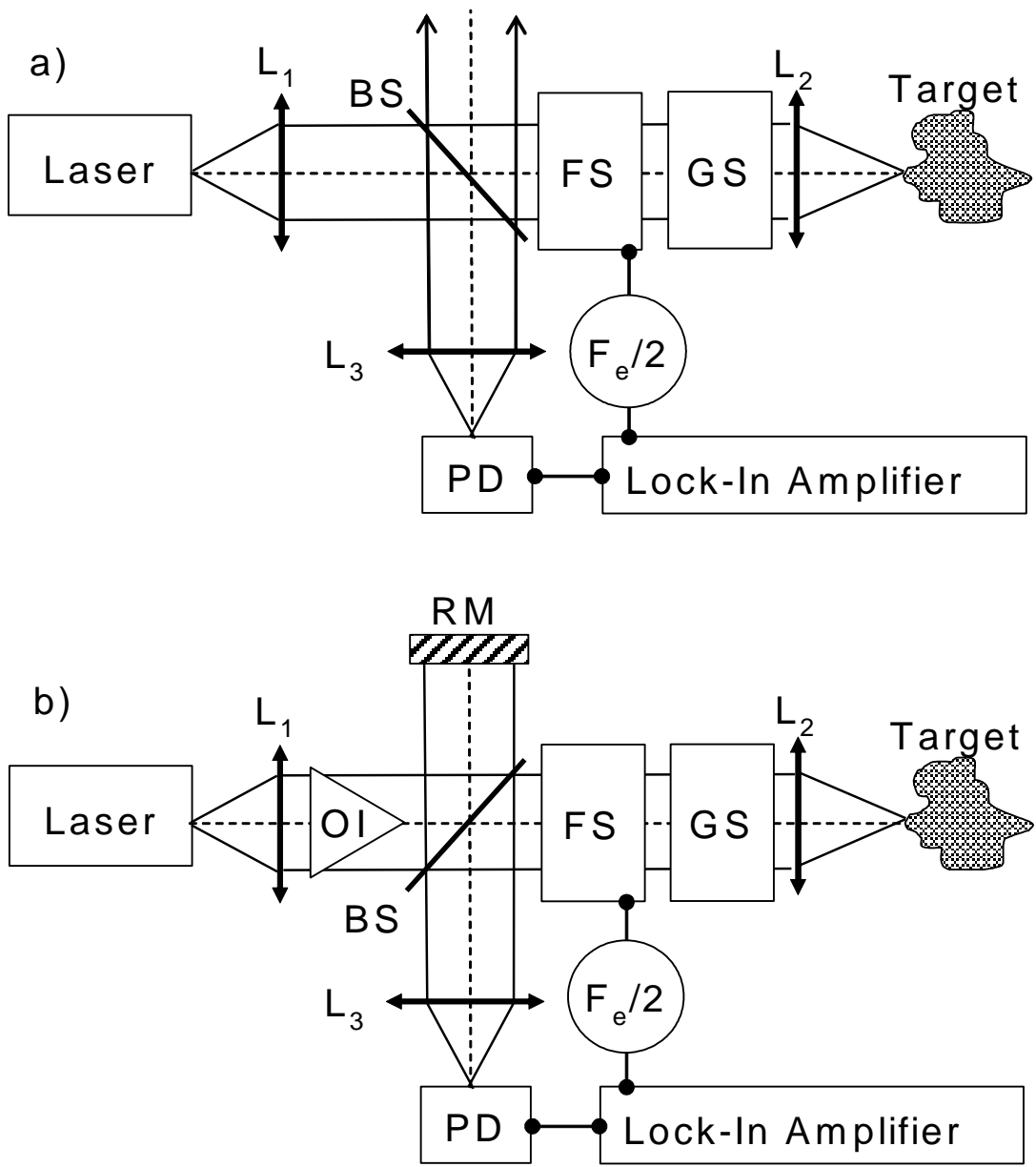

Fig. 1. Schematic diagrams of the autodyne interferometer setup (a) and of the heterodyne interferometer setup (b) for scanning microscopy. $L_1$, $L_2$ and $L_3$: Lenses, OI: Optical Isolator BS: Beam Splitter with a power reflectivity $R_{bs}$, GS: Galvanometric Scanner, RM: Reference Mirror with a unitary power reflectivity $R_{rm}=1$, FS Frequency Shifter with a round trip frequency-shift $F_e$, PD: Photodiode with a white noise spectrum. The lock-in amplifier is characterized by a bandwidth $\Delta F$ around the reference frequency $F_e$. The laser output power is characterized by $p_{out}$ (photons/s), the target is characterized by its effective reflectivity $R_e \ll 1$.

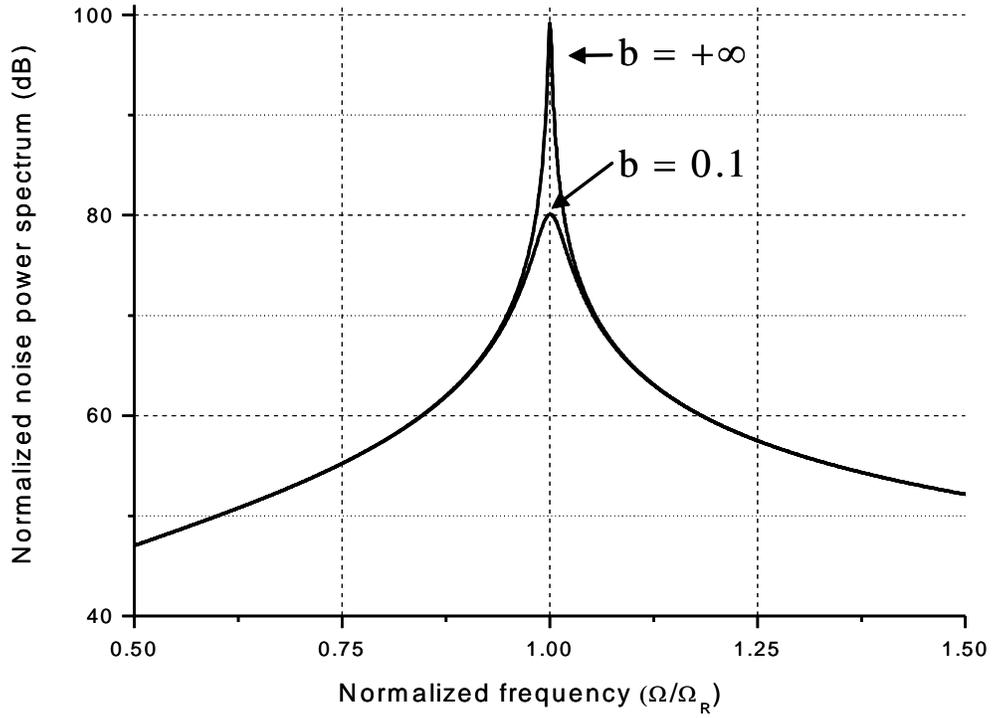

Fig. 2. Calculated normalized noise power spectra of the laser output power for two different values of the dimensionless parameter: $b = \gamma_b/\kappa$. The others laser dynamical parameters correspond to a conventional $Nd^{3+}$:$YVO_4$ microchip laser: $a = 2 \times 10^{-6}$, $a' = 2 \times 10^{-7}$, $\eta = 1/2$, $r = 5$ [19]. The 0 dB level corresponds to the shot noise level.

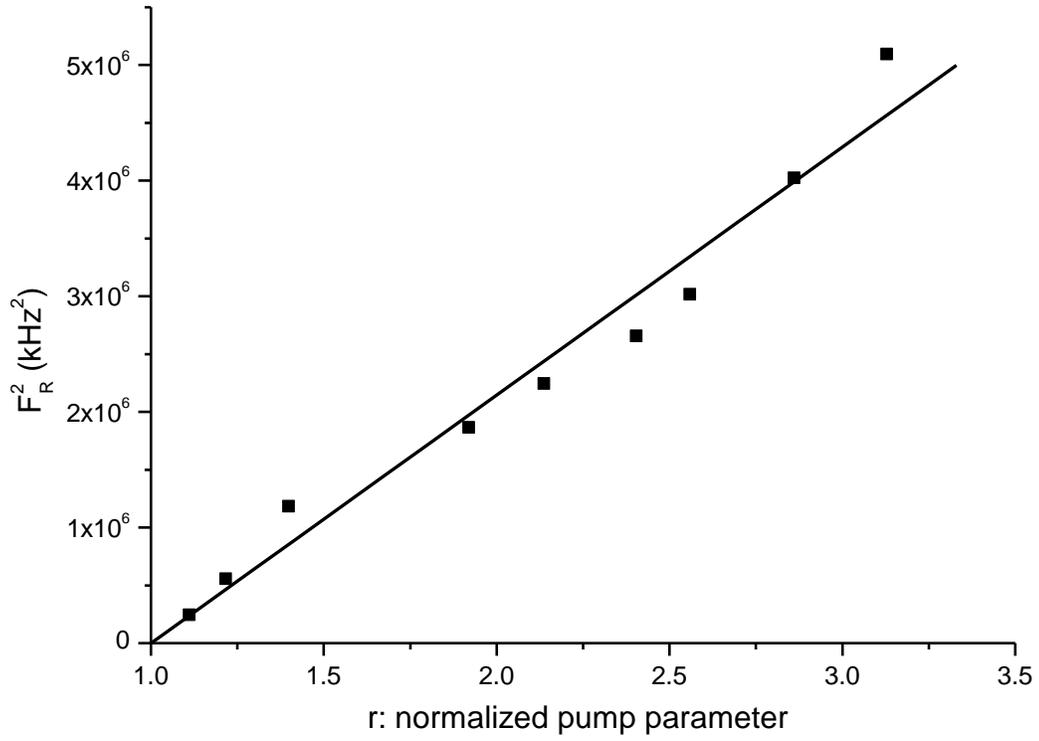

Fig. 3. Laser relaxation frequency ($F_R = \Omega_R/2\pi$) versus the normalized pumping parameter (r). Laser dynamical parameters: $\gamma_a = 3.3 \times 10^4 \, s^{-1}$ $\gamma_{a'} = 3.3 \times 10^3 \, s^{-1}$ and $\kappa = 2.1 \times 10^9 \, s^{-1}$.

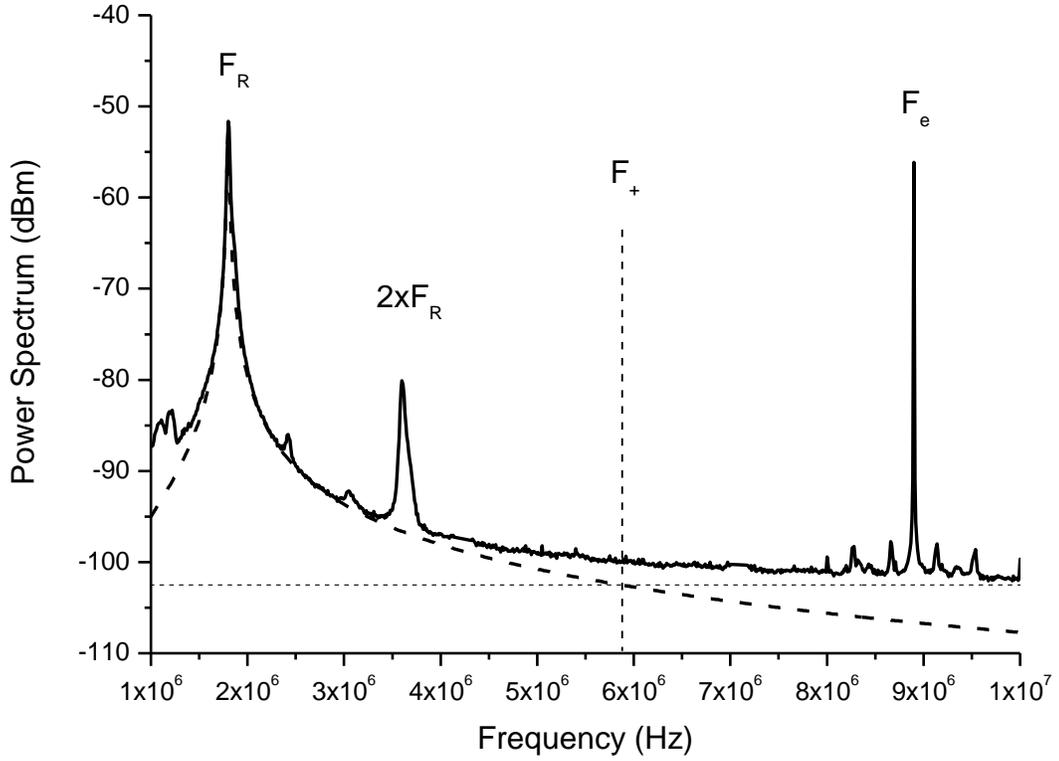

Fig. 4. Experimental (solid-line) and theoretical (dash-line) power spectra of the laser output power of a Nd:YVO$_4$ microchip laser with: $p_{out} = 2.14 \times 10^{17}$ photons /s ($P_{out} = 40$ mW at $\lambda = 1064$ nm ), $r = 2.6$, $\kappa = 2.1 \times 10^9$ s$^{-1}$, $\eta = 0.3$, $\gamma_a = 3.3 \times 10^4$ s$^{-1}$, $\gamma_{a'} = 3.3 \times 10^3$ s$^{-1}$, $\gamma_b = 3.3 \times 10^9$ s$^{-1}$. The experimental conditions are: $\Delta F = 500$ Hz ; $R_{bs} = 1/10$ ; $R_{load} = 50$ ohm . The horizontal dot line is the detection noise level.

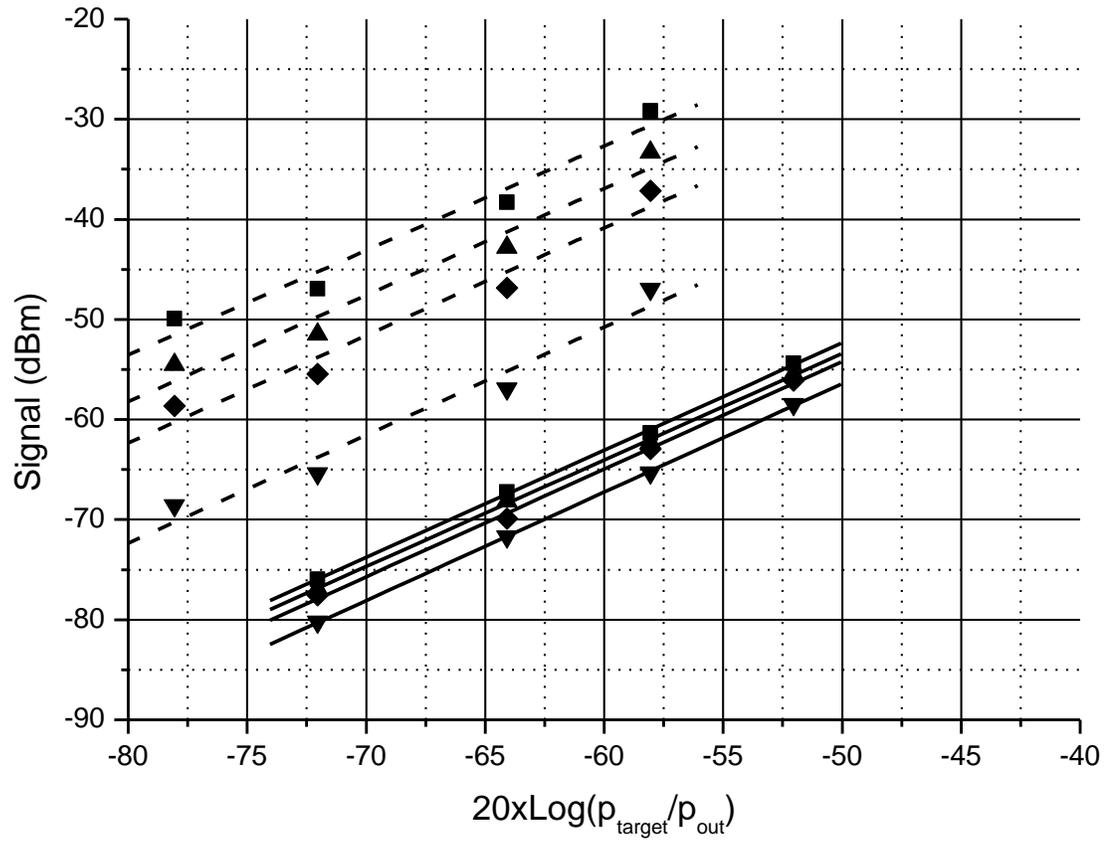

Fig. 5. Autodyne signal (dash-lines) and heterodyne signal (solid-lines) versus the laser power sent on the target ($p_{t\arg et}$), for different values of the shift frequency: (■) $F_e = 2.9 \times 10^6$ Hz , (▲) $F_e = 4.9 \times 10^6$ Hz , (◆) $F_e = 8.9 \times 10^6$ Hz , (▼) $F_e = 18.9 \times 10^6$ Hz .

**Table 1. Experimental ($\langle S_{LOFI}/S_{Heterodyne} \rangle$) and predicted ($G(Fe) \times \eta$) values of the LOFI signal gain. For each value of $F_e$, the average $\langle \ \rangle$ is made by using the signals obtained for the different values of $p_{target}$.**

| $F_e$ | 18.9 MHz | 8.9 MHz | 4.9 MHZ | 2.9 MHz |
|---|---|---|---|---|
| $\langle S_{LOFI}/S_{Heterodyne} \rangle$ | 6 ± 1 | 18 ± 3 | 28 ± 4 | 44 ± 5 |
| $G(Fe) \times \eta$ | 6 ± 2 | 13 ± 4 | 26 ± 7 | 62 ± 18 |

**Table 2.** Experimental ($\frac{\langle S_{LOFI} / N_{Laser} \rangle}{\langle S_{Heterodyne} / N_{Detection} \rangle}$) and predicted ($G(F_+) \times \eta$) values of the LOFI SNR gain. For each value of $p_{target}$, the average $\langle \rangle$ is made by using the results obtained for the different values of $F_e$.

| $P_{target}$ | 10 µW | 25 µW | 50 µW |
|---|---|---|---|
| $\langle S_{LOFI} / N_{Laser} \rangle$ | 233 ± 50 | 627 ± 133 | 1890 ± 431 |
| $\langle S_{Heterodyne} / N_{Detection} \rangle$ | 18 ± 3 | 47 ± 9 | 98 ± 16 |
| $\frac{\langle S_{LOFI} / N_{Laser} \rangle}{\langle S_{Heterodyne} / N_{Detection} \rangle}$ | 13 ± 4 | 13 ± 4 | 19 ± 5 |
| $G(F_+) \times \eta$ | 21 ± 6 | 21 ± 6 | 21 ± 6 |